\documentclass[11pt,a4paper,reqno]{amsart}
\usepackage{amsmath,amssymb,euscript,amsthm,amsfonts}
\usepackage{amsbsy}
\usepackage{amsopn}
\usepackage{graphicx}
\usepackage{multicol}

\begin{document}
\title[Fine-tune your smile: Correction to Hagan et al]{Fine-tune your smile\\{\footnotesize Correction to Hagan et al}}
\author[Jan Ob\l \'oj]{Jan Ob\l\'oj\smallskip\\
{\sc I\lowercase{mperial} C\lowercase{ollege} L\lowercase{ondon}}}
\thanks{Research supported by a Marie Curie Intra-European Fellowship within the $6^{th}$ European Community Framework Programme}
\address{\newline Imperial College London\newline Department of Mathematics\newline SW7 2AZ London, UK}
\email{j.obloj@imperial.ac.uk}
\date{January 2008. Please send comments to: \texttt{j.obloj@imperial.ac.uk}.}
\begin{abstract}
Using results derived in Berestycki \emph{et al.}~\cite{bbf04} we correct the celebrated formulae
of Hagan \emph{et al.}~\cite{hagan99,hagan02}. We derive explicitly the correct zero order term
in the expansion of the implied volatility in time to maturity. The new term is consistent as
$\beta\to 1$. Furthermore, numerical simulations show that it reduces or eliminates known pathologies
of the earlier formula.
\end{abstract}
\maketitle

We discuss here\footnote{
The present note originated from a seminar I co-organised at Imperial
College London in 2006/2007. I take the opportunity to thank Dr Mijatovi\'c and all the participants for fruitful discussions.
} the celebrated formulae of Hagan and Woodward \cite{hagan99} and Hagan \emph{et al.}~\cite{hagan02},
point out some inconsistencies and compare them with the results of Berestycki \emph{et al.}~\cite{bbf04}.
This leads to a new corrected version of the original formula derived in \cite{hagan02}.
This is a short technical paper and the reader is assumed to be familiar with the SABR model and asymptotic expansions of the implied volatility surface. We refer to Gatheral \cite[Chp.7]{gatheral06} for background.

Consider the following model for
the underlying (we take $r=0$ which should be interpreted as $S_t$ representing the forward value process)
\begin{equation}\label{eq:sabr}
\left\{%
\begin{array}{l}
 dS_t=\sigma_tS_t^\beta dW^1_t,\; S_0=s,\\
 d\sigma_t=\nu \sigma_t dW^2_t,\; \sigma_0=\alpha,\quad\textrm{where }d\langle W^1,W^2\rangle_t=\rho dt,\\
\end{array}%
\right.
\end{equation}
$\beta\in (0,1)$, $|\rho|\leq 1$, which is known as the SABR model. The model (\ref{eq:sabr})
is written in the pricing measure under which $W^1,W^2$ are correlated Brownian motions.
We take $\nu\geq 0$ and note that putting $\nu=0$ we obtain a local volatility model.
We refer interchangeably to strike $K$ or to $x=\ln(s/K)$. The implied volatility is denoted $I(x,\tau)$, i.e.\ a European call with strike
$K$ and time to maturity $\tau$ has the same price under the SABR model (\ref{eq:sabr}) and under the Black-Scholes model with volatility $I(x,\tau)$.

Consider now the Taylor expansion of the implied volatility surface $I(x,\tau)$ in time to maturity $\tau$ 
\begin{equation}\label{eq:series}
I(x,\tau)=I^0(x)\Big(1+I^1(x)\tau\Big)+O(\tau^2).
\end{equation}
Hagan \emph{et al.}~\cite{hagan02} were first to obtain
explicit expressions for $I^0$ and $I^1$. They used perturbation theory (combined with impressive
intuition). More recently, Berestycki \emph{et al.}~\cite{bbf04} treated
the subject in a rigorous analytical manner. They proved in particular that $I^0(x)=\lim_{\tau\to
0}I(x,\tau)$
is well defined and characterized it via a solution to eikonal equation (or via Varadhan's signed
geodesic distance). 
Direct, but somewhat tedious, computations allow us to compare their expression for $I^0$ with that of \cite{hagan02} and of \cite{hagan99} (case of $\nu=0$). The findings are summarised in Table
1.
\begin{table}
\begin{tabular}[h]{|l|ccc|l|}
\hline
& Hagan et al \cite{hagan02} && Berestycki et al \cite{bbf04}& \emph{Notation}\\
\hline
$I^0(0)$&$\alpha K^{\beta-1}$&=&$\alpha K^{\beta-1}$&\\
\hline
$\left.I^0(x)\right|_{\nu=0}$&
$\frac{x\alpha(1-\beta)}{s^{1-\beta}-K^{1-\beta}}$&=&
$\frac{x\alpha(1-\beta)}{s^{1-\beta}-K^{1-\beta}}$&\\ 
\hline
$\left.I^0(x)\right|_{\beta=1}$&$\nu x\Big/\ln\Big(\frac{\sqrt{1-2\rho z+z^2}+ z-\rho}{1-\rho}\Big)$&=&
$\nu x\Big/\ln\Big(\frac{\sqrt{1-2\rho z+z^2}+ z-\rho}{1-\rho}\Big)$
& $z=\frac{\nu x}{\alpha}$\\
\hline
$\left.I^0(x)\right|_{\beta<1}$&
$\nu x\frac{\zeta}{z}\Big/\ln\Big(\frac{\sqrt{1-2\rho \zeta+\zeta^2}+ \zeta-\rho}{1-\rho}\Big)$
&$\neq$&
$\nu x\Big/\ln\Big(\frac{\sqrt{1-2\rho z+z^2}+ z-\rho}{1-\rho}\Big)$
& \hspace*{-0.2cm}$
\left.%
\begin{array}{l}
 z=\frac{\nu}{\alpha}\frac{s^{1-\beta}-K^{1-\beta}}{1-\beta}\\
 \zeta=\frac{\nu}{\alpha}\frac{s-K}{(sK)^{\beta/2}}\\
 \end{array}%
\right.
$\\
\hline
\end{tabular}\bigskip
\caption{Comparison of $I^0$ term in \cite{hagan02} and \cite{bbf04}.}
\end{table}

The formula for $I^0(x)$ in the local volatility case $\nu=0$ agrees in both works and
appears to be correct. We note however that it differs from the formula in Hagan and Woodward
\cite{hagan99}. Numerically the difference is small but persistent. Unfortunately, it is the
formula in \cite{hagan99} which is often used, as in Bourgade and
Croissant \cite{bourgade06}.\smallskip

The formula for $I^0(x)$ when $\beta=1$ is the same in \cite{hagan02} and in \cite{bbf04}.
However the two papers differ\footnote{Note there is a typo in (6.5) in \cite{bbf04} which we corrected here.} when $\beta<1$ and we believe it is the formula of \cite{bbf04} which is correct\footnote{Strictly speaking Hagan \emph{et al.}~\cite{hagan02} do not claim to obtain the Taylor expansion in maturity, however their formula (A.65) is of the form \eqref{eq:series}.}
and should be used. This is natural as authors in \cite{bbf04} present an analytic treatment of the subject and characterize $I^0$ theoretically. Furthermore, the form of their formula we
derived here coincides with the general formula obtained by Labord\`ere \cite[Eq.~(5.1)]{labordere05}
(taken for the case of $0$-SABR). We give now two arguments in favor of the formula of \cite{bbf04}: theoretical and numerical.

From the theoretical point of view, the formula of Hagan \emph{et al.}~\cite{hagan02} has
an internal flaw: it is inconsistent as $\beta\to 1$. 
Indeed, $I^0_\beta(x)$ in the formula of \cite{hagan02} does \emph{not} converge to the known value for $I^0(x)_{\beta=1}$ as $\beta\to 1$.\footnote{However, when one uses the simplified formula (2.17a), which is (A.69c), instead of the general formula (A.65), the convergence issue (magically) disappears.} In contrast, as 
$$z=z(\beta)=\frac{\nu}{\alpha}\frac{s^{1-\beta}-K^{1-\beta}}{1-\beta}\longrightarrow \frac{\nu x}{\alpha},\quad
\textrm{when }\beta\to 1,$$
we have convergence using the formula of \cite{bbf04}.

The formula of Hagan \emph{et al.}~\cite{hagan02} is known to produce wrong prices in region of small
strikes for large maturities. It assigns a negative price to a structure with a positive payoff, or equivalently it
implies a negative probability density for the stock price process in some region. If one uses the formula we derive here the problem either appears in yet lower strikes or disappears completely.
To see this, we price the following structure in interest rates: long $3\%$ put and short $1\%$ put at three times the notional, whose payoff is a triangle, zero at zero and at $3\%$ and with a peak at $1\%$. We denote it $T(1)$. More generally, $T(K)$ is the following: long $(K+2)\%$ put and short $K\%$ put at $(K+2)/K$ times notional. The payoff of $T(K)$ is a triangle, zero at zero and at $(K+2)\%$, with a peak at $K\%$.

Let $I^0_H(x)$ and $I^0_B(x)$ be the formulae for $I^0$ from \cite{hagan02} and \cite{bbf04}
respectively. The implied volatility $I_H(x)$ in \cite[Eq.~(A.65)]{hagan02} is given via
\begin{equation}
\begin{split}
I_H(x)&=I_H^0(1+I_H^1(x)\tau),\\
I^1_H(x)&=\frac{(\beta-1)^2}{24}\frac{\alpha^2}{\big(sK\big)^{1-\beta}}+\frac{1}{4}\frac{\rho\nu\alpha\beta}{\big(sK\big)^{(1-\beta)/2}}
+\frac{2-3\rho^2}{24}\nu^2
\end{split}
\end{equation}
and we define the implied volatility $I_B(x)=I^0_B(x)(1+I^1_H(x)\tau)$, as we did not derive
an explicit formulae for $I^1_B$ from \cite{bbf04}. An explicit formula for $I^B_1$, albeit complicated, is derived in Labord\`ere \cite{labordere05}, who also argues that $I^1_H$ is a valid approximation.
\setlength{\columnsep}{0cm}
\begin{figure}[htbp]\label{fig:1}
\begin{multicols}{2}{
\includegraphics[height=5cm]{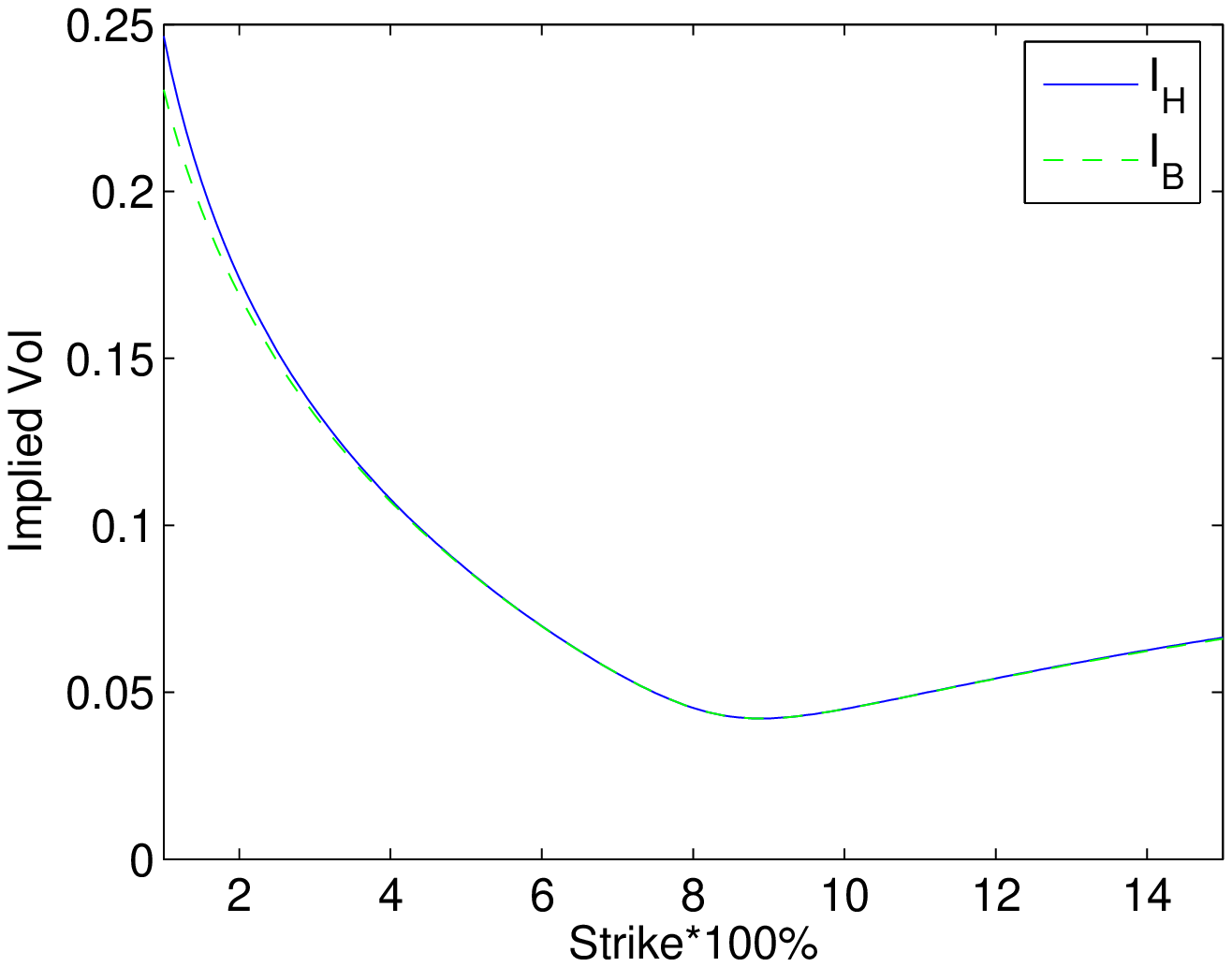}

\includegraphics[height=5cm]{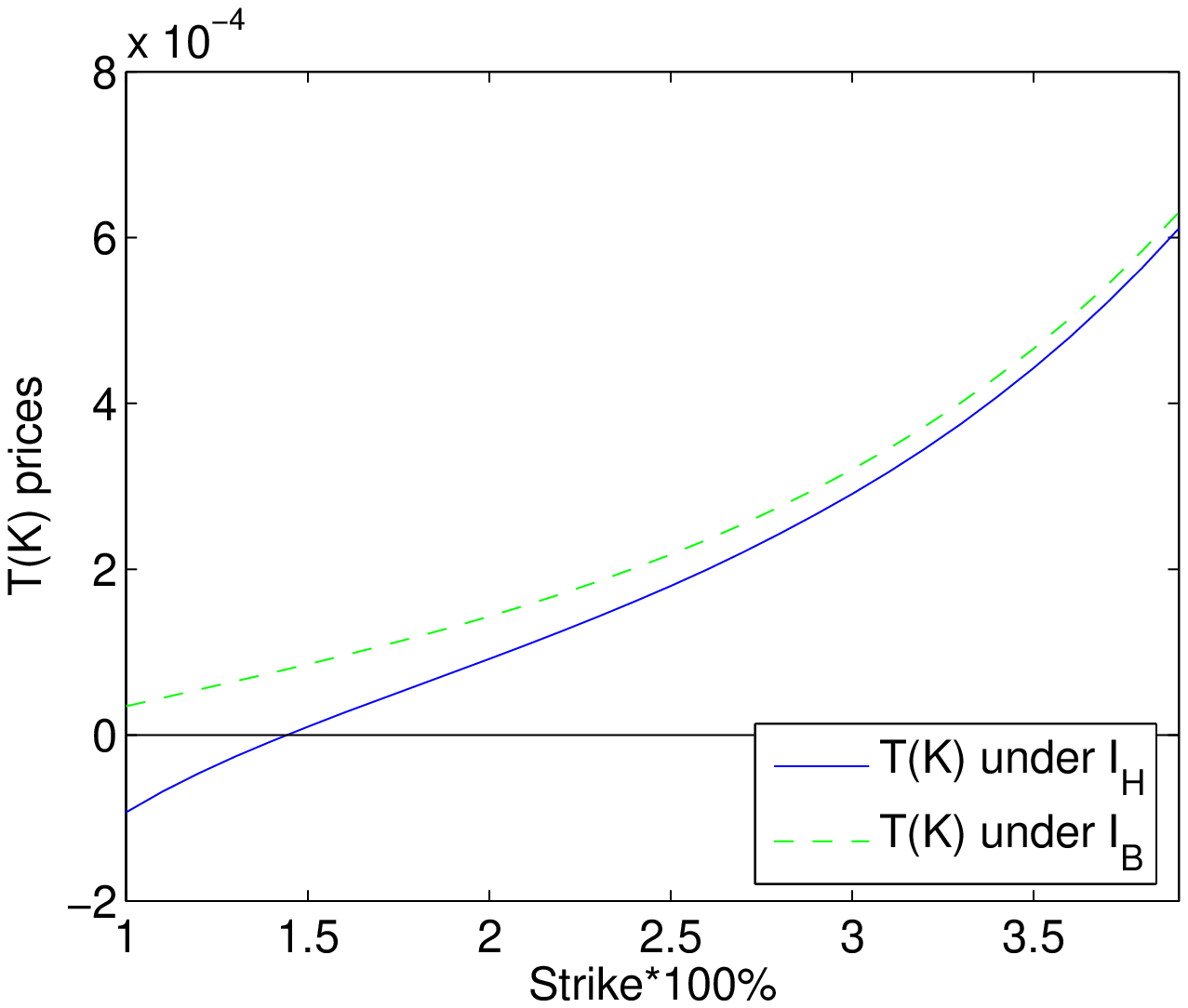}
}
\end{multicols}
\vspace*{-1cm}
\caption{\small Implied volatilities and $T(K)$ prices for maturity $\tau=15$\newline $\rho=-33\%$, $\beta=0.4$, $ATM=4.25\%$, $\nu=25\%$, $s=8.01\%$}
\end{figure}
\begin{figure}[htbp]\label{fig:2}
\begin{multicols}{2}{
\includegraphics[height=5cm]{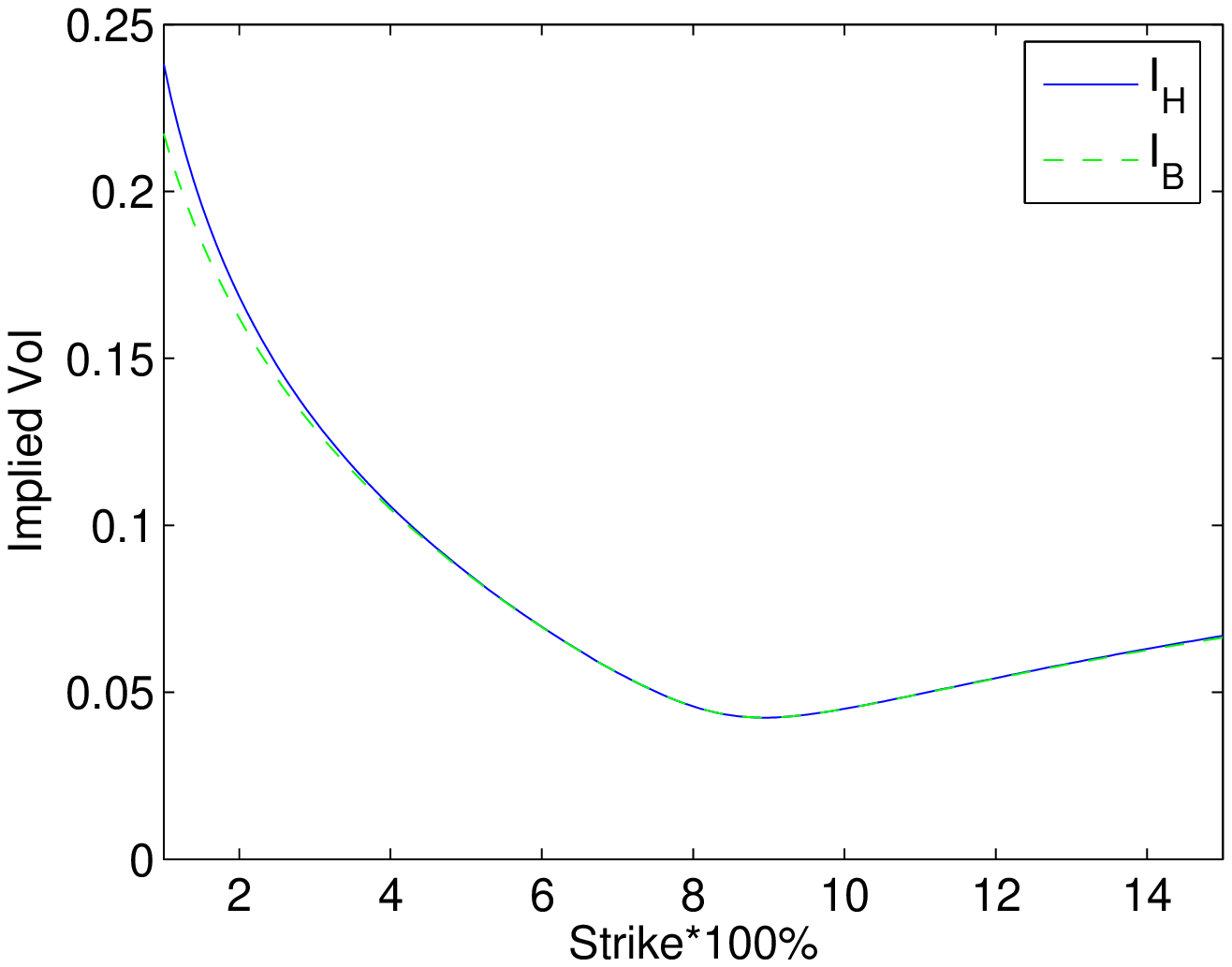}

\includegraphics[height=5cm]{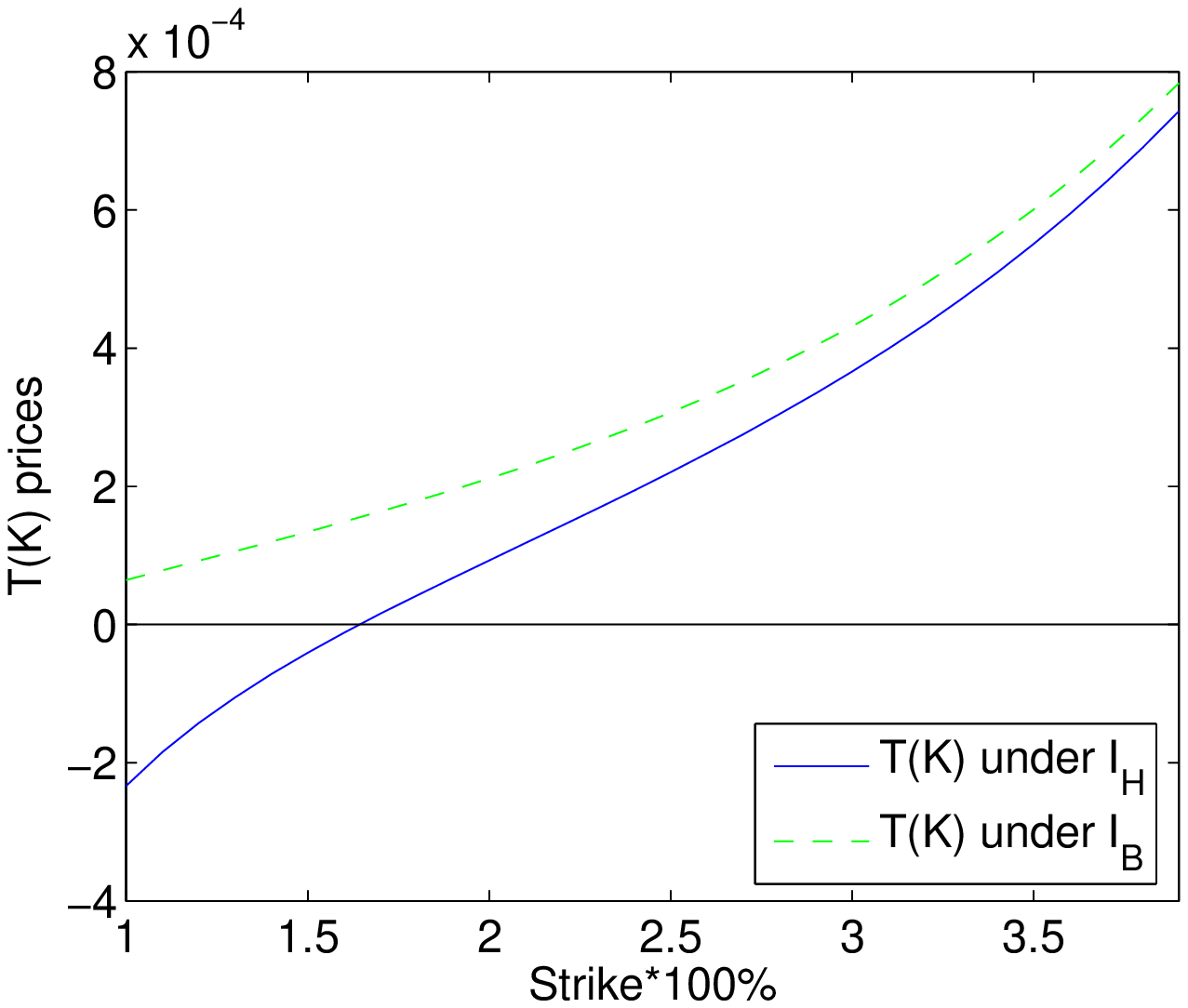}
}
\end{multicols}
\vspace*{-1cm}
\caption{\small Implied volatilities and $T(K)$ prices for maturity $\tau=20$\newline $\rho=-37\%$, $\beta=0.6$, $ATM=4.25\%$, $\nu=24.5\%$, $s=8.01\%$}
\end{figure}
Figures $1-2$ compare the prices of the structure
$T(K)$ derived using $I_H$ and $I_B$. The anomalies do not occur, or occur at lower
strikes, if we price using the implied volatility $I_B$ instead of $I_H$.
\newpage
\bibliographystyle{abbrv}
\bibliography{ivolbib}

\end{document}